# First-principles investigation of the physical properties of wide band gap hexagonal $AlPO_4$ compound for possible applications

A.S.M. Muhasin Reza, Md. Asif Afzal, S.H. Naqib*
Department of Physics, University of Rajshahi, Rajshahi 6205, Bangladesh
*Corresponding author; Email: salehnaqib@yahoo.com

## Abstract

$AlPO_4$ belongs to the berlinite quartz type family and is a ternary wide band gap semiconductor. The structure of $AlPO_4$ is hexagonal with space group 152 ($P3_12_1$). In this study, using density functional theory (DFT) method, we have investigated the bulk physical properties like structural, electronic band structure, elastic properties, thermal properties, optical properties and bonding features of $AlPO_4$ compound. The values of our optimized structural parameters are very close to previous results. Most of our calculations presented in this work are novel. The elastic constants indicate that $AlPO_4$ is mechanically stable and brittle in nature. The compound is hard and possesses low machinability index. $AlPO_4$ contains significant mechanical anisotropy. The charge density distribution, bond population analysis, Vickers hardness, thermomechanical and optical properties have been investigated theoretically for the first time. The values of Pugh's ratio and Poisson's ratio revealed the brittleness of the compounds associated with strong directional covalent bonds with a mixture of ionic contributions. From the bond population analysis the bonding character is also found to be mixed with ionic and covalent characters. Low values of Vickers hardness indicate the softness of the material. The electronic band structure calculations reveal clear insulating behavior with a band gap of ~6.0 eV. Band structure calculations were carried out without and with spin-orbit coupling (SOC) to explore possible topological signature. The energy dependent optical properties conform to the electronic band structure calculations. Major optical properties like dielectric functions, refractive index, photoconductivity, absorption coefficient, loss function and reflectivity are calculated and discussed in detail in this study. The compound is optically anisotropic. It is an efficient absorber and reflector of the ultraviolet light. Possible sectors of applications of $AlPO_4$ have been explored.

## 1. Introduction

In the past several years, research on the wide band gap semiconductors has led to major advances for device applications [1,2]. The merits of such material are for high-temperature electronics and short-wavelength optical applications. Ternary $AlPO_4$ is an important compound in this regard [3,4]. The thermal and chemical stability of $AlPO_4$ found at high temperatures and in abrasive environments also make them attractive for high-power operation under extreme environment [3]. This wide band gap ternary compound has been marked for its application in blue/green/UV light-emitting diodes, laser diodes, solar-blind UV detectors and high-power, high temperature electronics in the last decade [4]. On the other hand, possible applications of the $AlPO_4$ the in mechanical, engineering and thermal sectors remained largely unexplored.

From the X-ray diffraction (XRD) experiments [5], the crystal structures of $AlPO_4$ have been found in three different phases; first one is ($P\bar{l}$) phase at 6 GPa grown between 1000−1250°C, the 2nd is a $P2_1/c$ phase at 6−7 GPa grown at 1500°C, and the third one is $AlPO_4$ phase that was synthesized at a pressure region of 4−5 GPa which is stable in the berlinite (quartz form) phase. This phase was found to have a moganite-type structure. Moganite is a rare $SiO_2$ polymorph that has been recognized to have a widespread distribution in microcrystalline silica at the Earth's surface[6]. The phase transformation of crystalline to amorphous for the α-quartz-berlinite (α-$AlPO_4$) was firstly reported in 1990 in experiment using around 20 GPa pressure to a crystalline berlinite sample in a diamond anvil cell at 300 K. The structural parameters were found as follows: space group No. 152 ($P3_12_1$), a = b = 4.943 Å, and c = 10.947 Å [7]. Inorganic phosphates are usually large band gap materials which make them suitable as hosts for phosphors. The large band gap is associated with the large energy gap between the bonding and anti-bonding states in the phosphate groups [8].

To the best of our knowledge, there are no available experimental or theoretical studies on the bulk elastic, mechanical, acoustic, bonding, optical, lattice dynamical and thermo-mechanical properties of α-$AlPO_4$ yet. All these unexplored bulk physical properties are very important to understand this compound better and to unlock its full potential for applications in the engineering and optoelectronic sectors. The aim of this work is to fill these significant research gaps existing for α-$AlPO_4$ ($AlPO_4$ from now on). We have calculated the elastic constants and elastic moduli for the optimized crystal structure of $AlPO_4$. The significant mechanical performance indicators [9,10] like the brittleness/ductility, hardness, and machinability index have been evaluated. The elastic/mechanical/optical anisotropy parameters are estimated. Some thermomechanical parameters relevant to applications of $AlPO_4$ are computed as well. Most of the results presented in this study are novel.

The rest of the paper has been arranged as follows: the computational methodology is described in Section 2. The results of the calculations are presented and discussed in Section 3. Section 4 consists of the conclusions of this work.

## 2. Computational Scheme

In this research, for the calculations of the structural, mechanical, electronic, optical and thermal properties, we have used the CAmbridge Serial Total Energy Package (CASTEP) code [11] for the first-principles quantum mechanical calculations wherein the pseudopotential plane-waves (PP-PW) approach based on the density functional theory (DFT) [12] are adopted. DFT is thought of as the most precise scheme for solving the Kohn-Sham equations. We used the local density approximation LDA (CA-PZ) proposed by Ceperley and Alder and parameterized by Perdew and Zunder, named CA-PZ, as the exchange-correlation functional for the total energy calculations of the crystal unit cell [11,12]. The interactions between electrons and ion core are represented by a Vanderbilt-type ultrasoft pseudopotential [13]. The following setups are used for the DFT calculations: plane wave cut-off energy is set to 400 eV to ensure convergence, the Monkhorst-Pack scheme [14] is used for 15x15x6 mesh with 138 k points for sampling the first Brillouin zone (BZ). For the crystal structure optimization, the tolerances for self-consistent field is taken as $5.0 \times 10^{-7}$ eV/atom, energy as $5.0 \times 10^{-6}$ eV/ atom, maximum force as 0.01 eV/Å, maximum displacement is $5.0 \times 10^{-4}$ Å and a maximum stress of 0.02 GPa. All the DFT calculations were performed at the default temperature (0 K) and at default hydrostatic pressures zero (0) in GPa. The valence electron configurations of the elements are: Al-$3s^23p^1$, P-$3s^23p^3$ and O-$2s^22p^4$. The electronic wave functions and consequent charge density as well as the optimized (minimum total energy and internal forces) structural parameters of hexagonal $AlPO_4$ is calculated following the Broyden-Fletcher-Goldfarb-Shanno (BFGS) structural optimization. The total energy of each cell is calculated with the periodic boundary conditions. The hexagonal crystal structure of $AlPO_4$ has six independent elastic constants ($C_{11} = C_{22}$, $C_{12}$, $C_{13} = C_{23}$, $C_{33}$, $C_{44} = C_{55}$, and $C_{66}$). The elastic constants were obtained employing the ‘stress-strain’ method in the CASTEP program [15]. All the polycrystalline elastic moduli of $AlPO_4$, such as the bulk modulus (B), shear modulus (G) and Young’s modulus (E) were evaluated from those $C_{ij}$ by using the Voigt-Reuss-Hill (VRH) approximation[16,17]. The electronic band structure, total density of states (TDOS) and partial density of states (PDOS) are also calculated by using the optimized geometry of $AlPO_4$. Furthermore, we have used Quantum Espresso code to investigate the effect of spin-orbit coupling (SOC) on the electronic band structure of $AlPO_4$ within LDA.

The frequency-dependent optical spectra of the compound are derived from the complex dielectric function, $\varepsilon(\omega) = \varepsilon_1(\omega) + i\varepsilon_2(\omega)$. The imaginary part $\varepsilon_2(\omega)$ is obtained using the following expression [17]:

$$e_2(\omega) = \frac{2e^2\pi}{V\in_0}\sum_{k,v,c} \left| < y_k^c \middle| \hat{u}.\vec{r} \middle| y_k^n > \right|^2 \delta(E_k^c - E_k^n - E) \qquad (1)$$

In this expression, V represents the volume of the unit cell, ω is the (incident) photon angular frequency, $\varepsilon_0$ defines the dielectric constant of the free space, e is the charge of electron, **u** (unit vector) is defining the polarization of the incident electric field, **r** is the position vector, and $y_k^c$ and $y_k^n$ are the conduction and valence band wave functions for a given wave vector k, respectively. The real part of the dielectric function, $\varepsilon_1(\omega)$, has been evaluated via the Kramers-Kronig transformation, once $\varepsilon_2(\omega)$ has been computed. All the other optical parameters, such as complex refractive index n(ω), absorption coefficient α(ω), energy loss-function L(ω), reflectivity R(ω) and optical conductivity σ(ω) can be deduced from the known dielectric function [18–20]. The projection of the plane-wave (PW) states onto a linear combination of atomic orbitals (LCAO) basis sets is used to understand the bonding nature of $AlPO_4$, invoking Mulliken bond population analysis (MPA) and Hirshfeld bond population analysis (HPA) [21].

## 3. Results and Analysis

### *3.1. Structural Properties and Formation Energy*

As mentioned previously, we are interested in α-quartz berlinite (α-$AlPO_4$) type crystal structure [7] in this study. The schematic crystal structure is shown in Fig. 1.

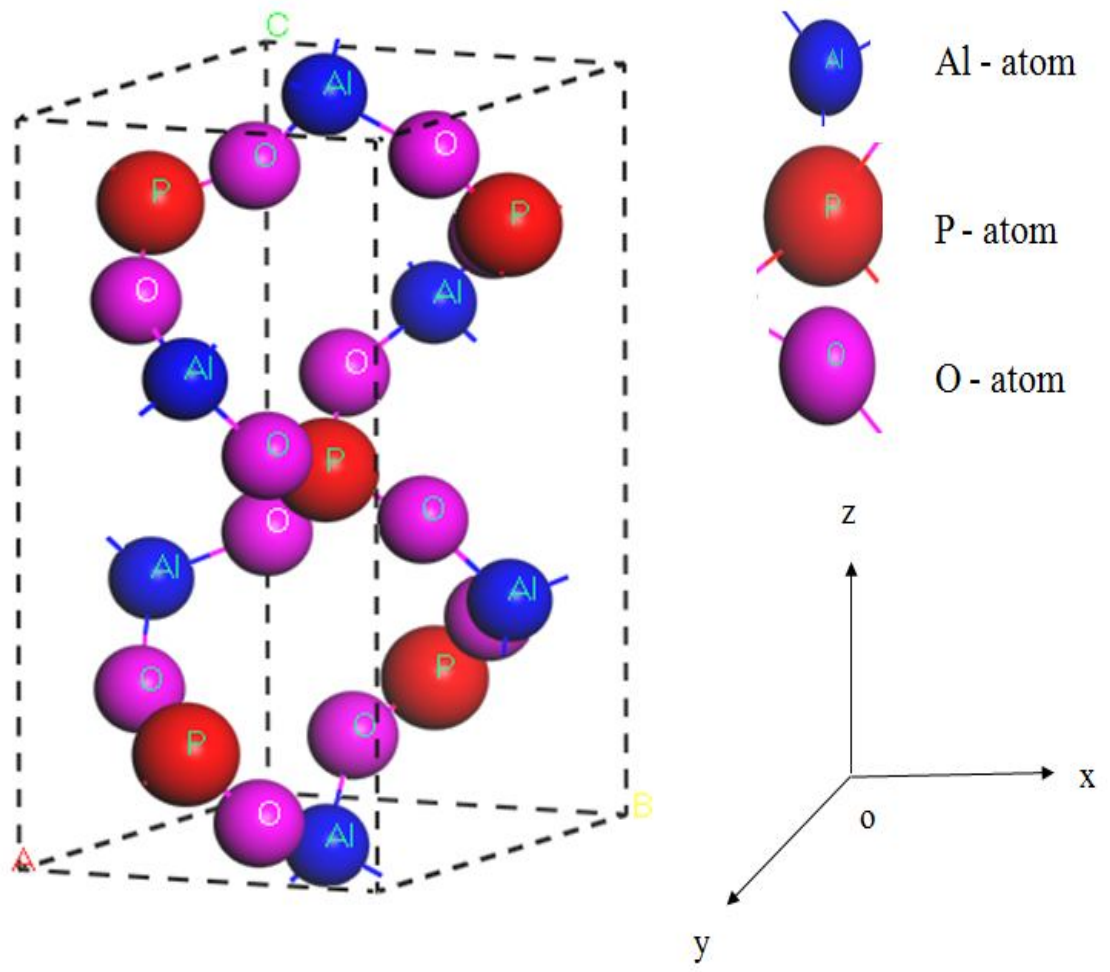

Figure 1: Schematic crystal structure of $AlPO_4$ unit cell. The crystallographic directions are shown.

The Wyckoff atomic positions for the elements in $AlPO_4$ are as follows: Al (0.467, 0, 0.333), P (0.467, 0, 0.833), O1 (0.422, 0.296, 0.397), O2 (0.409, 0.252, 0.886).

The first step in any of the ab-initio calculation is to determine the stable structure and finding the optimized geometry of the crystal. In this work, the atomic positions, c/a ratio, and a-, b-, and c-lattice parameters were relaxed. The calculated crystallographic lattice constants a (= b) and c for $AlPO_4$ are given in Table 1, together with previously reported values for comparison [8,22]. It is observed that the optimized structural parameters are very close to the experimental ones. This confirmed the reliability of our calculations. The hexagonal ratio c/a is 2.211, which is similar the previously reported experimental and theoretical values[8,22].

In order to investigate the phase stability of $AlPO_4$, its formation energy has been calculated. The energy of formation per atom is defined as:

$$E_{for}^{AlPO_4} = \frac{E_{total}^{AlPO_4} - xE_{solid}^{Al} - yE_{solid}^{P} - zE_{solid}^{O}}{(x+y+z)} \quad (2)$$

where x stands for number of Al atoms, y for P atoms and z for O atoms in the unit cell. We have x = 1, y = 1 and z = 4 for the $AlPO_4$ compound. $E_{total}^{AlPO_4}$, $E_{solid}^{Al}$, $E_{solid}^{P}$ and $E_{solid}^{O}$ represents the total energy values for the $AlPO_4$ phases and Al, P and O atoms in the most stable solid forms, respectively. Our calculated results for the formation energy are also displayed in Table 1. This calculation gives the formation energy of -8.52 eV/atom for the hexagonal $AlPO_4$. One can see that the value of the formation energy is negative, which implies that, from the thermodynamic point of view, the materialization of $AlPO_4$ is exothermic and this compound is chemically stable, in covenant with the experiment.

Table 1: Calculated and lattice constants a = b and c, c/a ratio, equilibrium cell volume, number of formula units in the unit cell Z and total number of formula units N in the conventional cell of hexagonal $AlPO_4$.

| Compound | a = b (Å) | c (Å) | c/a | Volume $V_0$ (Å$^3$) | Formation energy (eV) | Z | N | Ref. |
|---|---|---|---|---|---|---|---|---|
| $AlPO_4$ | 4.942 | 10.947 | 2.215 | -- | | -- | -- | [7]Theo. |
| | 4.941 | 10.940 | 2.214 | -- | | -- | -- | [22]Exp. |
| | 4.890 | 10.816 | 2.211 | 224.045 | -8.52 | 3 | 18 | This work |

*3.2. Elastic and Mechanical Properties*

3.2.1. The Elastic Constants

The mechanical properties such as elastic stability, stiffness, brittleness, ductility, and elastic anisotropy of a solid can be obtained from the elastic constants. All these parameters are important to select a compound for engineering applications. The results for the five independent elastic constants ($C_{11}$, $C_{12}$, $C_{13}$, $C_{33}$, and $C_{44} = C_{55}$) and one dependent elastic constant $C_{66} = (\frac{C_{11}-C_{12}}{2})$ applicable for $AlPO_4$ in the hexagonal geometry are listed in Table 2. The elastic constants satisfy the Born stability[23] criteria: $C_{11} > 0$, $(C_{11}–C_{12}) > 0$, $C_{44} > 0$, $(C_{11} + C_{12}) C_{33} –2C_{13} > 0$, that confirms that the compound is mechanically stable. For this compound $C_{33} > C_{11}$ (Table 2), hence, the atomic bonding is stronger along c-axis than that of along a-axis. Since $C_{33}$ are larger than $C_{44}$ the linear compression along the crystallographic c-axis is difficult compared to the shear deformation.

Table 2: Calculated elastic constants $C_{ij}$ (in GPa) for $AlPO_4$ compound.

| Compound | $C_{11} = (C_{22})$ | $C_{12}$ | $C_{13}$ (= $C_{23}$ ) | $C_{33}$ | $C_{44}$ (= $C_{55}$) | $C_{66}$ | Ref. |
|---|---|---|---|---|---|---|---|
| $AlPO_4$ | 64.99 | 0.839 | 5.359 | 79.11 | 47.376 | 32.075 | This work |

Overall, the values of $C_{ij}$ are quite small ($C_{12}$, in particular) signifying that the atomic bonds in the compound under study are weak. $C_{12}$ is the parameter for dilation on compression, i.e., it indicates that an axial stress results in a strain along a perpendicular axis.

3.2.2. The Elastic Moduli

To comprehend the bulk mechanical properties with clarity, the elastic moduli (Y, B, and G), Pugh ratio (G/B or B/G) and Poisson's ratio (σ) are calculated. The elastic modului have been calculated from the elastic constants, $C_{ij}$, using the Voigt-Reuss-Hill (VRH) method [16,17], which are shown in Table 3. For pure solids the fracture strength is proportional to their bulk modulus (B) and lattice parameter. The bulk modulus indicates the resistance to the volume change against external stress, indicating the average bond strength. The rigidity against plastic deformation or resistance to change in the bonding angle is proportional to the shear modulus (G). The lower value of G than B predicts plasticity will dominate its mechanical strength. The bulk modulus is proportional to the cohesive energy and represents, on average, the material's opposition to bond rupture. The larger value of G manifests the more pronounced directional bonding between atoms [24]. For a polycrystalline material, in the Voigt scheme, the aggregate strain is considered uniform in line with the external strain whereas in the Reuss approximation scheme, aggregate stress is considered uniform in line with the external stress. In the Hill scheme an arithmetic average of the Voigt and Reuss approaches are taken.

To calculate the bulk mechanical properties, the following equations are used [16,17]:

$$B = \frac{1}{2}(B_V + B_R) \tag{3}$$

and

$$G = \frac{1}{2}(G_V + G_R) \tag{4}$$

where

$$B_V = \frac{1}{9}\{2(C_{11} + C_{12}) + 4C_{13} + C_{33}\} \tag{5}$$

and

$$G_V = \frac{1}{30}(C_{11} + C_{12} + 2C_{33} - 4C_{13} + 12C_{44} + 12C_{66}) \tag{6}$$

where $B_V$ and $G_V$ are the B and G in terms of the Voigt approximation. And

$$B_R = \frac{(C_{11}+C_{12})C_{33}-2C_{12}^2}{(C_{11}+C_{12}+2C_{33}-4C_{13})} \tag{7}$$

$$G_R = \frac{\frac{5}{2}\left[\left((C_{11}+C_{12})C_{33}-2C_{12}^2\right)^2\right]C_{55}C_{66}}{\left[3B_V C_{55}C_{66}+\left(\left((C_{11}+C_{12})C_{33}-2C_{12}^2\right)^2(C_{55}+C_{66})\right)\right]} \tag{8}$$

$B_R$ and $G_R$ are the B and G in terms of the Reuss approximation. We have used the standard relationships [25] to calculate the Young's modulus (Y) and Poisson's ratio ($\sigma$) as follows:

$$Y = \frac{9BG}{(3B+G)} \tag{9}$$

$$s = \frac{(3B-2G)}{2(3B+G)} \tag{10}$$

The stiffness of a solid is estimated by the value of its Young modulus (Y). The higher value of Y indicates that the material is stiffer.

Table 3: The calculated values of bulk modulus (B), shear modulus (G) and Young's modulus (Y) (all in GPa) of $AlPO_4$.

| Compound | $B_R$ | $B_V$ | $B_H = B$ | $G_R$ | $G_V$ | $G_H = G$ | B/G | Y | Ref. |
|---|---|---|---|---|---|---|---|---|---|
| $AlPO_4$ | 25.419 | 25.800 | 25.610 | 34.225 | 38.534 | 36.379 | 0.704 | 74.054 | This work |

The values of all the elastic moduli are moderate. This is a consequence of the low values of $C_{ij}$.

For the understanding of mechanical characteristics of a material the bulk elastic constants are necessary. In engineering, condensed matter physics, geophysics, materials science, and chemistry research fields [26,27] these parameters play important roles. The stability, plasticity, stiffness, brittleness, ductility, chemical bonding, thermal characteristics and anisotropy of material can all be calculated using the elastic constants and moduli.

### 3.2.3. Poisson's Ratio

The Poisson's ratio (σ), is linked to the material's stability against the shearing stress which gives knowledge regarding its chemical bonding and brittle or ductile failure mode. Poisson's ratio is determined by the ratio of transverse strain to longitudinal strain for a given tensile stress. Moreover, from the Poisson ratio, the brittleness or ductility of the solid can be understood. For brittle materials, Poisson ratio is found to be small, whereas, for ductile systems, the typical value of the Poisson ratio (σ) is found to be around 0.30 or higher [28]. The computed Poisson's ratio σ is extremely low, 0.018, for $AlPO_4$ which is much less than the critical value (0.26) [29] indicating its highly brittle nature. Such low value of the Poisson's ratio suggests that the chemical bonding in $AlPO_4$ is highly directional. A crystalline solid is always stable under either central force or non-central force. A material will be stabilized by central force if its Poisson's ratio lies between 0.25 and 0.50, otherwise it will be stabilized by non-central force [30]. Thus, in case of $AlPO_4$, the non-central type force stabilizes its hexagonal crystal structure.

It is worth noting that, very low Poisson's ratio solids are useful for a number of practical applications. For example, such solids are used for packing materials, medical knee pads, foot wears, and for materials requiring long-lasting shock absorption [10].

### 3.2.4. The Machinability Index

The machinability properties of a material are one of the important parameter in the manufacturing fields. On the machinability index the cutting conditions, drilling rates, feed rate, tool wear, and machining time are dependent. The machinability property of a solid also reflects the dry-lubricity and plasticity (resistance to plastic deformation) behaviors[31]. If the machinability of a compound increases, the Plasticity also increases. The machinability index, $\mu_m$ of a material is defined as[31]:

$$\mu_m = \frac{B}{C_{44}} \tag{11}$$

The machinability and dry-lubricity will be better if the bulk modulus is higher with the lower shear resistance. The calculated value of $\mu_m$ is disclosed in Table 5. The value of $\mu_m$ of our compound $AlPO_4$ is 0.54. This value is quite low and it is a consequence of highly brittle nature of the compound.

### 3.2.5. The Pugh Ratio

The Pugh ratio (B/G) is closely related to the brittleness and ductility of a material. For brittle (ductile) material, the value of Pugh ratio is higher (lesser) than the critical value of 0.57. The value of B/G of $AlPO_4$ is 0.704, which confirms its brittleness [31]. The prediction from the Pugh ratio matches perfectly with the prediction from the Poisson's ratio.

3.2.6. The Hardness

Hardness is a fundamental parameter of a solid in the engineering field to design different devices. Using the elastic moduli we can calculate the hardness value. We have estimated here two types of hardness; micro-hardness ($H_{micro}$) and macro-hardness ($H_{macro}$). $H_{micro}$ hardness is linked to the Young modulus and Poisson's ratio, while $H_{macro}$ is estimated from shear and bulk moduli, sometimes called the Chen hardness. Theoretical description underlying both the hardness formulae can be found elsewhere [32]. Therefore, deviation of results for different calculations of hardness seems to be reasonable. It is noted that the values of $H_{micro}$ and $H_{macro}$ are 7.857 GPa and 11.687 GPa, respectively, which suggest that the compound under study is moderately hard. The hardness values of $AlPO_4$ are estimated from the elastic moduli using the following macroscopic relationships found in existing literature [24,33]:

$$H_1 = 0.0963B \tag{12}$$
$$H_2 = 0.0607Y \tag{13}$$
$$H_3 = 0.1475G \tag{14}$$
$$H_4 = 0.0635Y \tag{15}$$
$$H_5 = -2.899+0.1769G \tag{16}$$
$$H_6 = \frac{(1-2\sigma)B}{6(1+\sigma)} \tag{17}$$
$$H_{micro} = \frac{(1-2\sigma)Y}{6(1+\sigma)} \tag{18}$$
$$H_{macro} = 2\left[\left(\frac{G}{B}\right)^2 G\right]^{0.585} - 3 \tag{19}$$

Table 4: The calculated values of hardness (GPa) of $AlPO_4$.

| Compound | $H_1$ | $H_2$ | $H_3$ | $H_4$ | $H_5$ | $H_6$ | $H_{micro}$ | $H_{macro}$ | Remarks |
|---|---|---|---|---|---|---|---|---|---|
| $AlPO_4$ | 2.46 | 4.49 | 5.36 | 4.70 | 3.53 | 4.04 | 7.85 | 11.68 | This work |

3.2.7. The Cauchy Pressure

The Cauchy pressure, $C_p$ is another important mechanical parameter for materials. It provides us with information regarding the bonding nature of solids. For hexagonal system, the $C_p$ is calculated for different directions by the following formulae: $C_{Px} = (C_{13} - C_{44})$ and $C_{Py} = (C_{12} - C_{66})$. The value of $C_{Px}$ and $C_{Py}$ of $AlPO_4$ are -42.017 GPa and -31.236 GPa, respectively. A negative $C_P$ indicates the covalent bonding and a positive values indicates the ionic bonding of the compound [34,35]. Furthermore, a positive Cauchy pressure suggests intrinsic ductility and a negative Cauchy pressure suggests brittleness of the material. Pettifor's rule [36] states that the positive $C_p$ implies non-directional bonding and negative $C_p$ reflects angular bonding in a solid.

3.2.8. The Tetragonal Shear Modulus

The parameter, called as the tetragonal shear modulus, is a measure of crystal stiffness (the resistance to shear deformation by a shear stress applied in the (110) plane in the $[1\bar{1}0]$ direction) and is given by:

$$C^{¢} = \frac{(C_{11} - C_{12})}{2} \quad (20)$$

This parameter is related to the velocity of low-frequency sound in a solid. A positive value of the tetragonal shear modulus is an indication of dynamical stability of a crystal structure. Calculated value of of $C^{¢}$ of $AlPO_4$ is positive and given in Table 5.

3.2.9. The Kleinman Parameter

The stability of a solid against stretching and bending can be measured by an elastic parameter called the Kleinman parameter (ζ) which is also known as internal strain parameter. The ζ of $AlPO_4$ has been calculated using following equation [36]:

$$X = \frac{C_{11} + 8C_{12}}{7C_{11} + 2C_{12}} \quad (21)$$

The calculated value is displayed in Table 5. Small value of the Kleinman parameter of $AlPO_4$ implies that the mechanical strength of this compound is dominated by bond bending contributions.

Table 5: Calculated values of Poisson's ratio (σ), machinability index ($\mu_m$), Pugh's ratio (B/G), Cauchy pressure $C_p$ (GPa), tetragonal shear modulus C′ (GPa) and Kleinman parameter (ξ) of $AlPO_4$ compound.

| Compound | σ | $\mu_m$ | B/G | $C_p$ | C′ | ξ | Ref. |
|---|---|---|---|---|---|---|---|
| $AlPO_4$ | 0.018 | 0.540 | 0.704 | -46.531 | 32.075 | 0.157 | This work |

All the elastic parameters shown in Tables 2-5 are novel and should be used as references for future studies.

3.2.10. Elastic Anisotropy

Elastic/mechanical anisotropy is an important factor in materials engineering because of the performance of the material under external loading (stress/strain) depends on its directional bonding characteristics. In practice, elastic anisotropy controls a large number of physical characteristics of solids, like alignment/misalignment of grain boundaries, movement of micro-cracks in materials, phonon dynamics, defect mobility, development of plastic deformation in crystals and its mechanical durability. Generally, anisotropic properties dominate in the covalent crystals and isotropic properties dominate in the solids with metallic bonding [24]. We summarize the expressions used to calculate various anisotropy indices of hexagonal $AlPO_4$ below [18,37] :

The Zener anisotropy factor A is defined as:

$$A = \frac{2C_{44}}{(C_{11} - C_{12})} \quad (22)$$

This is a measure of overall anisotropy in shear. The direction dependent shear anisotropy factors are given below.

For the {100} shear planes between the ⟨011⟩ and ⟨010⟩ directions, it is:

$$A_1 = \frac{4C_{44}}{(C_{11}+C_{33}-2C_{13})} \quad (23)$$

for the {010} shear plane between ⟨101⟩ and ⟨001⟩ directions, it is:

$$A_2 = \frac{4C_{55}}{(C_{22}+C_{33}-2C_{23})} \quad (24)$$

and for the {001} shear planes between ⟨110⟩ and ⟨010⟩ directions, it is:

$$A_3 = \frac{4C_{66}}{(C_{11}+C_{22}-2C_{12})} \quad (25)$$

The calculated shear anisotropy factors are summarized in Table 6. For elastically isotropic solids $A_1 = A_2 = A_3 = 1$; any deviation from 1 defines the degree of anisotropy. The computed values of $A_1$ (= $A_2$) and $A_3$ are 1.42 and 0.99, respectively. This suggests that the compound is almost completely isotropic with respect to the {001} shear plane.

The universal anisotropy index $A^U$, equivalent Zener anisotropy measure $A^{eq}$, anisotropy in shear $A^G$ (or $A^C$), and anisotropy in compressibility $A^B$ of crystals with any symmetry can be estimated from the following relations [24,38]:

$$A^u = 5\frac{G_V}{G_R} + \frac{B_V}{B_R} - 6 \geq 0 \quad (26)$$

$$A^{eq} = \left(1 + \frac{5}{12}A^u\right) + \sqrt{(1 + \frac{5}{12}A^u)^2 - 1} \quad (27)$$

$$A^G = \frac{(G_V - G_R)}{2G_H} \quad (28)$$

$$A^B = \frac{(B_V - B_R)}{(B_V + B_R)} \quad (29)$$

$A^U$ is the first anisotropy parameter among all other anisotropy indices which accounts for both shear and bulk contributions. $A_U$ is zero for isotropic crystals, whereas values other than zero imply anisotropy. The calculated value of $A^U$ is enclosed in Table 6. The value of $A^U$ of $AlPO_4$ is 0.644 which indicates its anisotropic nature.

For isotropic materials, the equivalent Zener anisotropy measure $A^{eq}$ is equal to one. The calculated value of $A^{eq}$ of our compound is 2.043 (Table 6) which predicts the anisotropic nature. The indices $A^G$ and $A^B$ represent the percentage anisotropy in shear and compressibility, respectively. $A^G = A^B = 0$ imply perfect isotropy, while $A^G = A^B = 1$ (100%) represent highest possible anisotropy. The values of anisotropy in shear $A^G$ or $A^C$ is 5.92% and the anisotropy in compressibility $A^B$ is very small, 0.761%. The larger value of $A^G$ than $A^B$ indicates much lower level of anisotropy in compressibility than in shear.

The universal log-Euclidean index ($A^L$) is given by[31]:

$$A^L = \sqrt{\left[ln\frac{B_V}{B_R}\right]^2 + 5\left[ln\frac{C_{44}^V}{C_{44}^R}\right]^2} \quad (30)$$

Here, $C_{44}^{V}$ and $C_{44}^{R}$ refer to the Voigt and Reuss values of $C_{44}$, respectively, with-

$$C_{44}^{R} = \frac{5}{3}\left\{\frac{C_{44}(C_{11}-C_{12})}{3(C_{11}-C_{12})+4C_{44}}\right\} \quad (31)$$

and

$$C_{44}^{V} = C_{44}^{R} + \frac{3}{5}\left\{\frac{(C_{11}-C_{12}-2C_{44})^2}{3(C_{11}-C_{12})+4C_{44}}\right\} \quad (32)$$

The limiting value of $A^L$ ranges from 0 to 10.27 [39]. For isotropic crystals the values of the universal log-Euclidean index is zero and it increases with the degree of anisotropy. The estimated value of $A^L$ is tabulated in Table 6. A large value of this parameter suggests layered feature of the crystal structure. The value of $A^L$ (6.17) of $AlPO_4$ compound is large, thus it is expected to exhibit layered structural characteristics, including a possibility of exfoliation.

Table 6: The Zener anisotropy factor A, shear anisotropy factors ($A_1$, $A_2$ and $A_3$), the universal anisotropy index $A^U$, equivalent Zener anisotropy measure $A^{eq}$, anisotropy in compressibility $A^B$ (%), anisotropy in shear $A^G$ (%) and the universal log-Euclidean index $A^L$ for $AlPO_4$.

| Compound | A | $A_1$ | $A_2$ | $A_3$ | $A^U$ | $A^{eq}$ | $A^B$ | $A^G$ | $A^L$ | Ref. |
|---|---|---|---|---|---|---|---|---|---|---|
| $AlPO_4$ | 1.476 | 1.420 | 1.420 | 0.999 | 0.644 | 2.043 | 0.761 | 5.92 | 6.17 | This work |

Moreover, the linear compressibilities of $AlPO_4$ along a- and c-axis ($\beta_a$ and $\beta_c$) were evaluated from [24]:

$$\beta_a = \frac{(C_{33}-C_{13})}{(C_{11}+C_{12})C_{33}-2(C_{13})^2} \quad (33)$$

$$\beta_c = \frac{(C_{11}+C_{12}-2C_{13})}{(C_{11}+C_{12})C_{33}-2(C_{13})^2} \quad (34)$$

The ratio between the coefficients, $\left(\frac{\beta_c}{\beta_a}\right) = 1$, for isotropic compressibility, and any other values indicates the anisotropy in the compressibility. The calculated value of $\left(\frac{\beta_c}{\beta_a}\right)$ is shown in Table 7 indicating anisotropy.

Furthermore, the uniaxial bulk modulus of solids can be determined through the single crystal elastic constants. The bulk modulus in the relaxed state and uniaxial bulk moduli along the a-, b-, and c-axis and anisotropies in the bulk modulus of $AlPO_4$ are determined from the following formulae[24]:

$$\beta_{relax} = \frac{\grave{U}}{(1+a+b)^2} \quad (35)$$

$$B_a = a\frac{dP}{da} = \frac{\grave{U}}{(1+a+b)} \quad (36)$$

$$B_b = a\frac{dP}{db} = \frac{B_a}{\alpha} \quad (37)$$

$$B_c = c\frac{dP}{dc} = \frac{B_a}{\beta} \quad (38)$$

and

$$A_{B_a} = \frac{B_a}{B_b} = \alpha \quad (39)$$

$$A_{B_c} = \frac{B_c}{B_b} = \frac{\alpha}{\beta} \quad (40)$$

with

$$\boldsymbol{\grave{U}} = C_{11} + 2C_{12}\alpha + C_{22}\alpha^2 + 2C_{13}\beta + C_{33}\beta^2 + 2C_{23}\alpha\beta \quad (41)$$

$$\alpha = \frac{(C_{11}-C_{12})(C_{33}-C_{13})-(C_{23}-C_{13})(C_{11}-C_{13})}{(C_{33}-C_{13})(C_{22}-C_{12})-(C_{13}-C_{23})(C_{12}-C_{23})} \quad (42)$$

and

$$\beta = \frac{(C_{22}-C_{12})(C_{11}-C_{13})-(C_{11}-C_{12})(C_{23}-C_{12})}{(C_{22}-C_{12})(C_{33}-C_{13})-(C_{12}-C_{23})(C_{13}-C_{23})} \quad (43)$$

$A_{B_a}$ and $A_{B_c}$ show the bulk anisotropy along the a- and c-axis with respect to b-axis, respectively. The calculated values are listed in Table 7. The bulk modulus anisotropy of the $AlPO_4$ is in the order, $B_c > B_a$ (= $B_b$). The ratios $A_{B_a}$ and $A_{B_c}$ are also given in Table 7.

Table 7: The bulk modulus ($B_{relax}$ in GPa), bulk modulus along a-, b- and c- axis ($B_a$, $B_b$ and $B_c$ in GPa), anisotropy in bulk modulus along a- and c- axes ($A_{Ba}$ and $A_{Bc}$), linear compressibility along a- and c- axes ($\beta_a$ and $\beta_c$ in $TPa^{-1}$) and the ratio of linear compressibility $\alpha = \left(\frac{\beta_c}{\beta_a}\right)$ for $AlPO_4$.

| Compound | $B_{relax}$ | $B_a$ | $B_b$ | $B_c$ | $A_{B_a}$ | $A_{B_c}$ | $\beta_a$ | $\beta_c$ | α | Ref. |
|---|---|---|---|---|---|---|---|---|---|---|
| $AlPO_4$ | 40.019 | 109.932 | 109.932 | 404.260 | 1 | 1.333 | 0.014 | 0.743 | 0.743 | This work |

3.2.11. The 2D and 3D Plots of Elastic Parameters

The 2D and 3D plots of the directional dependences of Young's modulus (Y), linear compressibility (β), Shear modulus (G) and Poisson ratio (σ) of $AlPO_4$ are also studied with the help of the ELATE generated graphs [40].

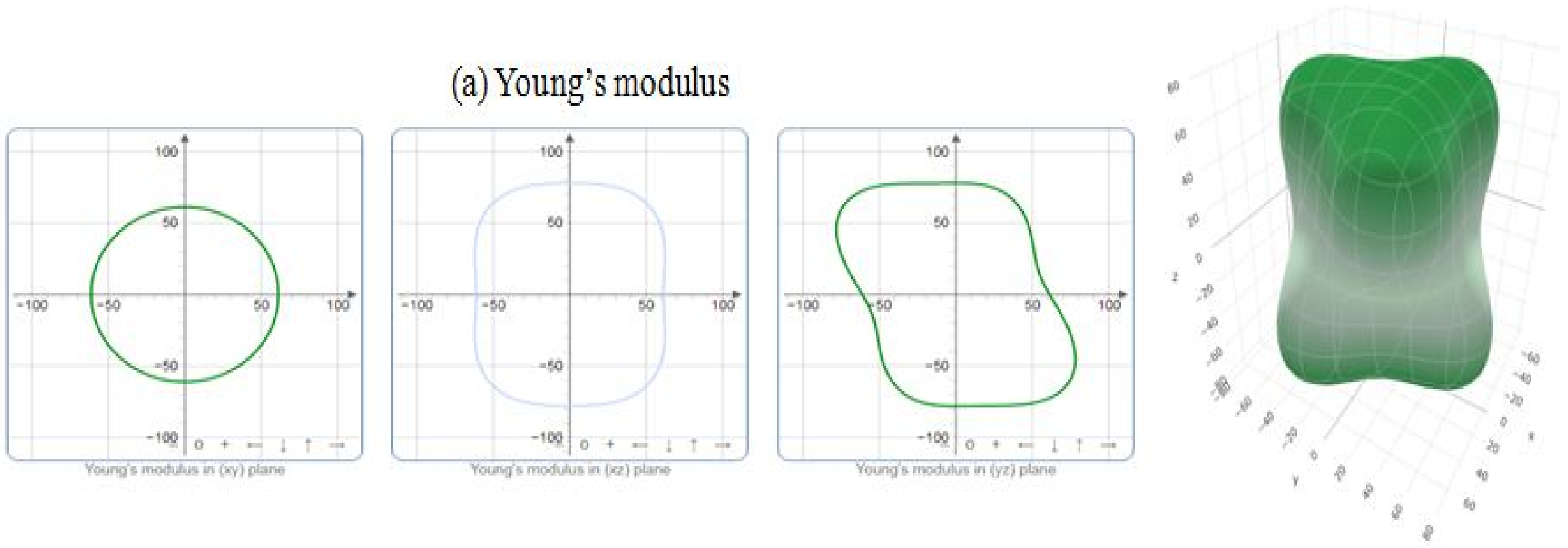

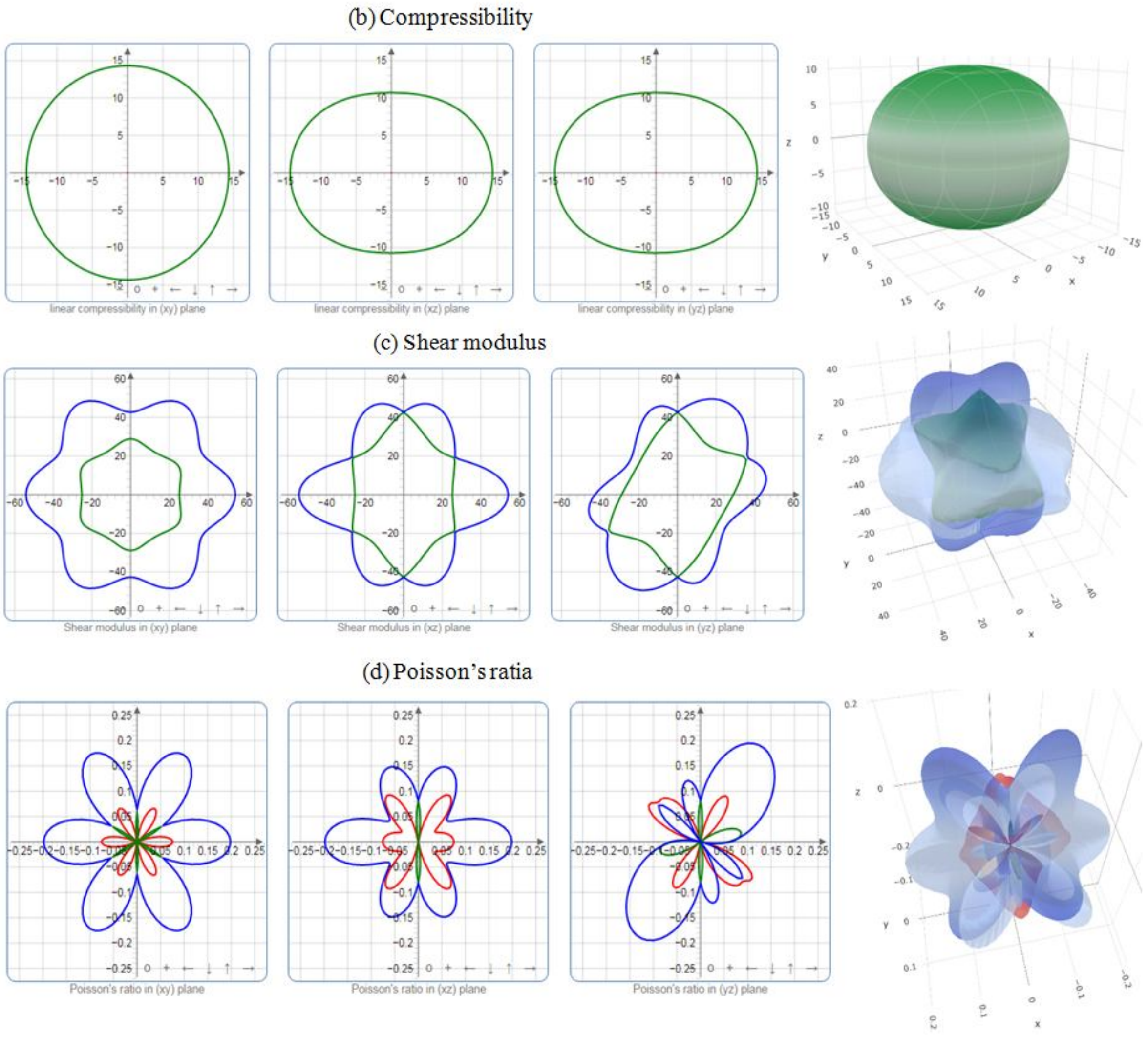

Figure 2: Directional dependences of (a) Young's modulus (Y), (b) compressibility (β), (c) shear modulus (G) and (d) Poisson's ratio (σ) of $AlPO_4$.

For this study we have used the elastic stiffness matrices. Uniform circular 2D and spherical 3D plots are the manifestations of the isotropic nature of crystals. The deviations from these ideal shapes show the degree of anisotropy. The 2D projection on the xy- , xz-, and yz-planes, along with the 3D view of Y, β, G and σ for $AlPO_4$ are seen in Figure 2. The green curves represent the minimum and the blue curves represent the maximum points of the relevant parameters. It is evident that two parameters are isotropic in the xy-plane, while they are anisotropic in other planes. Both 2D and 3D plots show the anisotropy in the following order: $\beta < Y < G < \sigma$. ELATE software also calculated a quantitative analysis of the minimum and maximum values of all the parameters involved and their ratios, as listed in Table 8.

Table 8: The minimum and maximum values of Young's modulus Y (GPa), compressibility β ($TPa^{-1}$), shear modulus G (GPa), Poisson's ratio σ and their ratios for $AlPO_4$ in the ground state.

| Y | | | β | | | G | | | σ | | |
|---|---|---|---|---|---|---|---|---|---|---|---|
| $Y_{min}$ | $Y_{max}$ | $A_Y$ | $\beta_{min}$ | $\beta_{max}$ | $A_\beta$ | $G_{min}$ | $G_{max}$ | $A_G$ | $\nu_{min}$ | $\nu_{max}$ | $A_\sigma$ |
| 56.47 | 96.53 | 1.70 | 10.70 | 14.31 | 1.33 | 25.27 | 54.17 | 2.14 | -0.12 | 0.22 | 1.86 |

Highest level of anisotropy is found in the shear modulus.

### *3.3. Electronic Band structure and TDOS & PDOS Features*

#### 3.3.1. Band Structure

The electronic energy versus momentum curve constitutes the electronic band structure of a material. The electronic transport properties, atomic bonding, superconductivity and optical response of a system are determined largely by the electronic band structure. The transport properties are controlled by the energy bands close to the Fermi level. For the modeling of nanostructures and electronic devices band topologies and effective masses of charge carriers are important which can be calculated from the electronic band structure [24,41,42]. We have calculated the energy band structure of hexagonal $AlPO_4$ along the high symmetry directions in the first Brillouin zone using the equilibrium lattice parameters as shown in Figure 3.

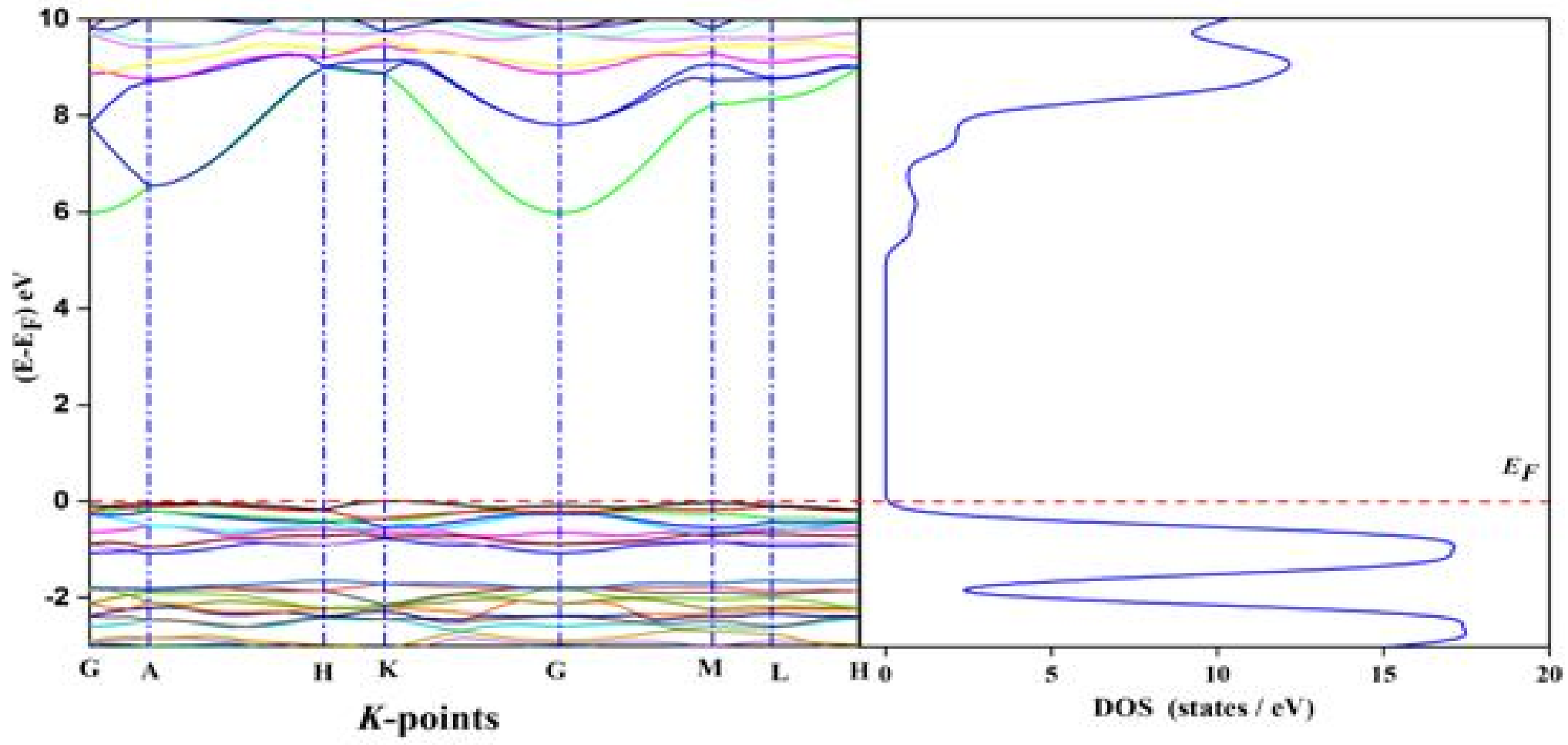

Figure 3: The electronic band structures of $AlPO_4$ in the ground state (obtained using CASTEP). The dashed horizontal line marks the Fermi energy (set to 0 eV). The panel in the right shows the energy density of states associated with the bands.

The non-metallic nature is observed from the non-overlapping of the conduction bands and valence bands at the Fermi level $E_F$ (horizontal dashed line placed at zero energy). The compound $AlPO_4$ is a large band gap semiconductor (insulator). The value of the indirect band gap between valance band and conduction band is around 5.97 eV. In the past several years, research on the wide-band-gap semiconductors have led to major advances which now make them viable for device applications. The valence band just below the Fermi level is derived mainly from the O-2p electronic states. The conduction band above the Fermi energy originates from the Al-3s and Al-3p electronic states. The O-2p band in the valence band is characterized by very large hole effective mass, while the Al-3s and Al-3p bands in the conduction region are highly hybridized and are characterized by small electron effective mass.

To explore the effect of SOC, electronic band structure of hexagonal $AlPO_4$ was revisited using the Quantum Espresso code. The electronic energy dispersion curves with and without SOC are displayed below (Fig. 4).

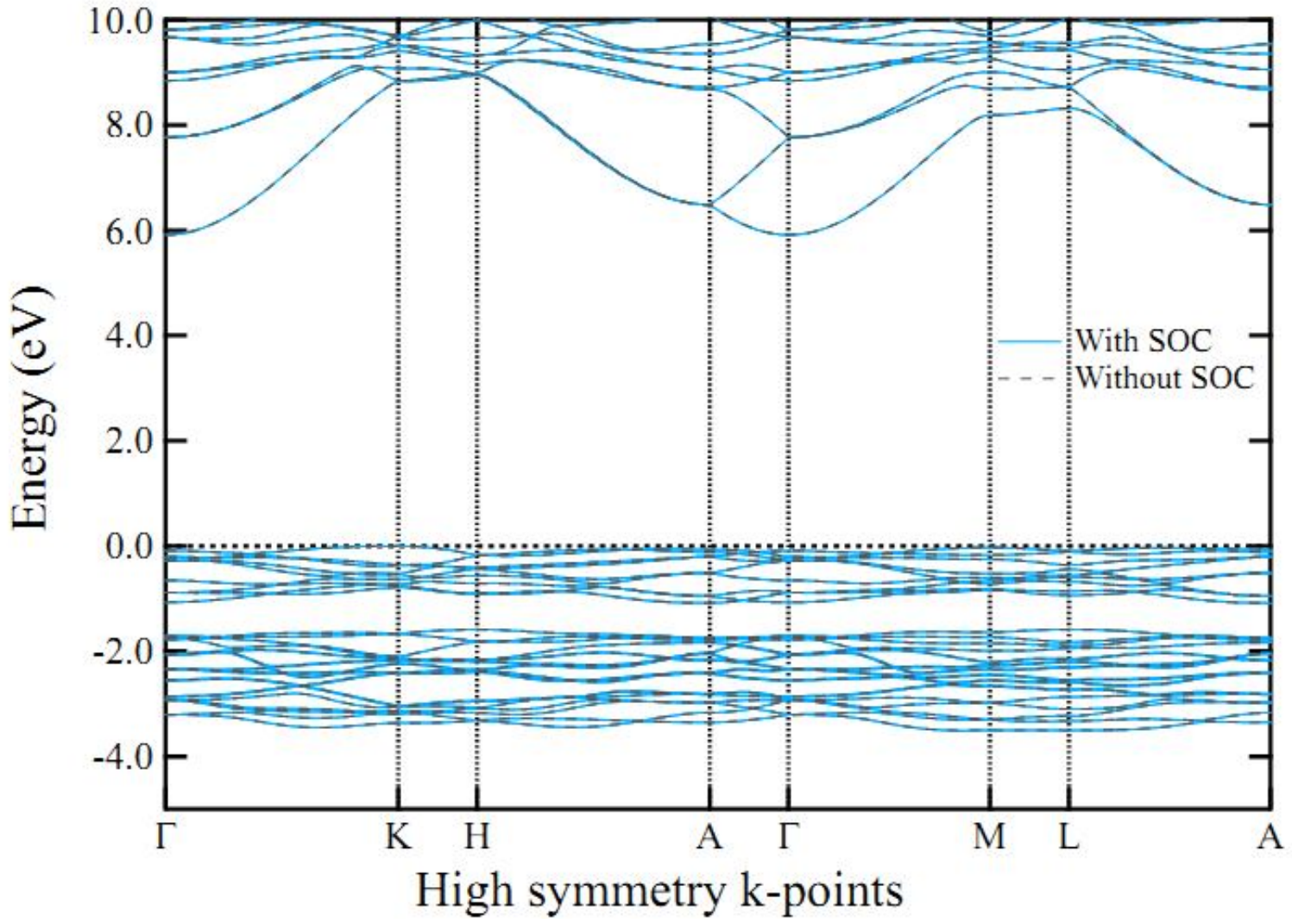

Figure 4: The electronic band structures of $AlPO_4$ in the ground state with and without SOC (obtained using Quantum Espresso). The dashed horizontal line marks the Fermi energy (set to 0 eV).

It is observed from Fig. 4 that the effect of the SOC on the electronic band structure is quite weak in $AlPO_4$ in the hexagonal structure. The value of the band gap is ~6.0 eV in both the cases.

3.3.2. TDOS & PDOS Features

For understanding of the bonding nature between atoms and the contribution of electrons to electronic conductivity, thermal conductivity, optical properties, information about the total density of states (TDOS) and the partial electronic density of states (PDOS) of a material are necessary. The DOS also determines various physical quantities, like electronic contribution to the heat capacity and spin paramagnetic susceptibility of a metals, which are directly related to the electronic density of states at the Fermi level [18,42]. Figure 5 shows the TDOS and PDOS of $AlPO_4$ compound. The vertical dashed line at zero energy level represents the Fermi energy.

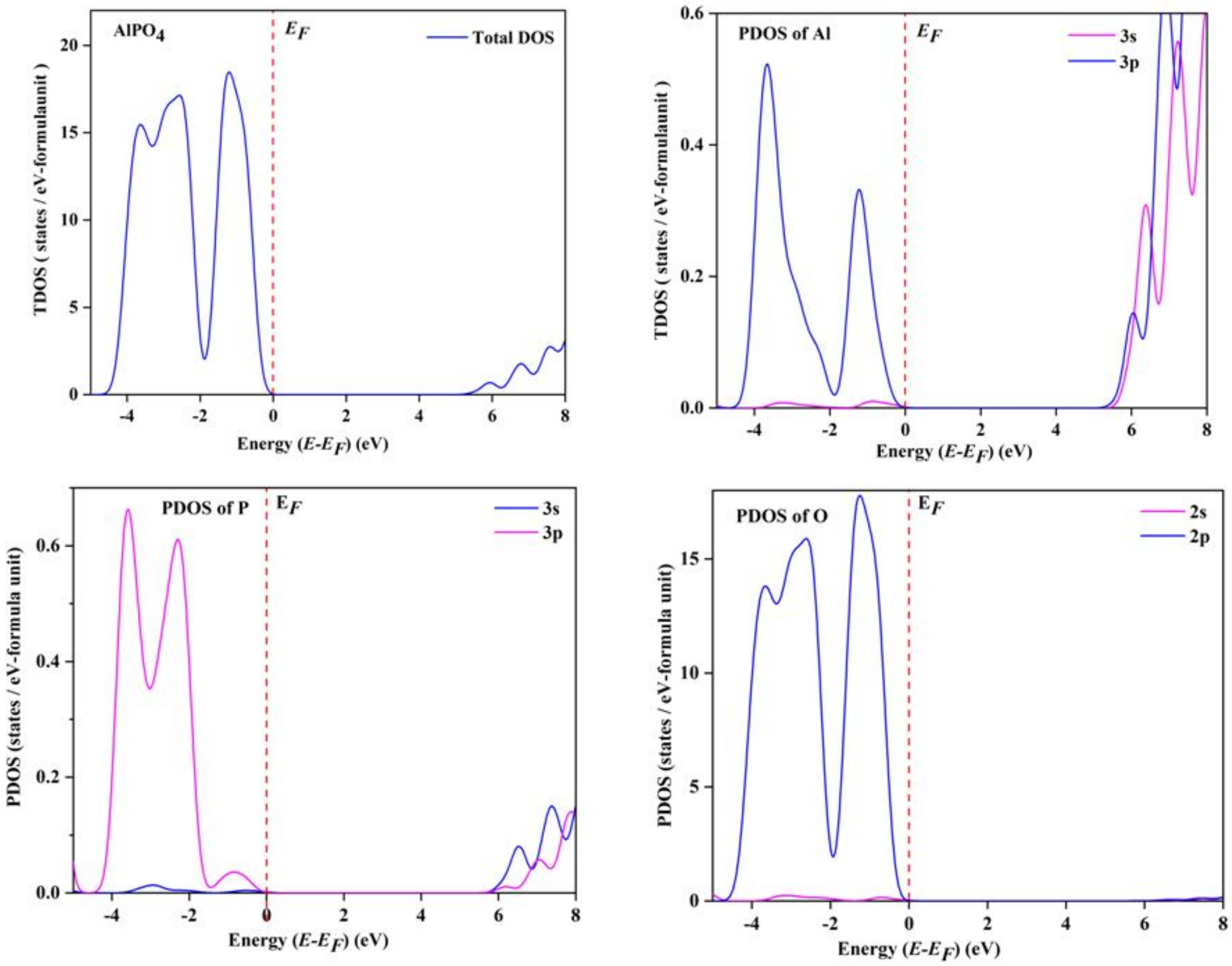

Figure 5: Total density of states (TDOS) and partial density of states (PDOS) of $AlPO_4$ as a function of energy. Fermi level is set to 0 eV.

There is no available electronic state at the Fermi level showing the non-metallic nature of the compound. TDOS is very high in the valence band close to the Fermi energy. This results from the localized nature of the O-2p electronic orbitals. There is energy hybridization among the electronic orbitals of O, P and Al atoms in the valence band. Such hybridization can lead to covalent bondings in a solid. The main contribution to the conduction band comes from the Al-3s and Al-3p electronic states.

*3.4. Bonding Analysis*

The chemical bonding nature and hardness of $AlPO_4$ are further studied by applying the Mulliken atomic population (MAP), effective valence charge (EVC) and bond overlap population (BOP) analyses [21]. From these calculations we saw that the total electronic charge for O atoms is more than any other atoms of our compound. Total charge of P is higher than the total charge of Al. The atomic Mulliken charges of Al, P, and O atoms are 2.11e, 2.45e and -1.14e, respectively. All the Mulliken atomic charges are deviated from their formal ionic charge which expected for the purely ionic state. Thus, covalent bondings are predicted in $AlPO_4$. The positive Mulliken atomic charges refer that the electrons are given away from the atom and the negative atomic Mulliken charges refer that the electrons are taken by the atoms. From atomic Mulliken charge calculations it is found that an ionic contribution persists between O-Al and O-P atoms. There is also covalent bonding between O-Al, Al-Al, Al-P, and P-P atoms because the number of electrons transferred to O atoms is not the same as the number of electrons transferred from any other atoms. For this, the bonding nature in $AlPO_4$ is expected to be mixed with both ionic and covalent features.

The effective valence charge (EVC) is calculated by the difference between formal ionic charge and Mulliken charge of an atom in a crystal [31,43]. If the value of EVC is zero, it indicates an ionic bond and a non-zero value indicates covalent bonding existing in the compound. The high values of EVC signify the high level of covalency in chemical bonds. All these information including the charge spilling parameter are disclosed in Table 9.

Sometimes the Hirshfeld population analysis (HPA) gives us a more accurate estimate of the EVC [34,43]. We have estimated the Hirshfeld charge of the material, which are also given in Table 9. The atomic charges of each atom calculated from the HPA are much smaller than the atomic charges calculated from the MPA. Nevertheless, both the HPA and the MPA calculations indicate that electrons are transferred to O from Al and P atoms.

Table 9: Charge Spilling parameter (%), orbital charge (electron), atomic Mulliken charge (electron), effective valance and Hirshfeld charge (electron) of $AlPO_4$ compound.

| Charge Spilling (%) | Atoms | Mulliken atomic populating | | | | | Mulliken charge | Formal ionic charge | Effective valence (Mulliken) | Hirshfeld charge | Effective valence (Hirshfeld) |
|---|---|---|---|---|---|---|---|---|---|---|---|
| | | s | p | d | f | total | | | | | |
| 1.43 | O | 1.83 | 5.31 | 0.00 | 0.00 | 7.14 | -1.14 | -1 | -0.14 | -0.25 | 0.75 |
| | O | 1.83 | 5.31 | 0.00 | 0.00 | 7.14 | -1.14 | -1 | -0.14 | -0.25 | 0.75 |
| | Al | 0.34 | 0.55 | 0.00 | 0.00 | 0.89 | 2.11 | 3 | 0.89 | 0.48 | 2.52 |
| | Al | 0.34 | 0.55 | 0.00 | 0.00 | 0.89 | 2.11 | 3 | 0.89 | 0.48 | 2.52 |
| | P | 0.80 | 1.75 | 0.00 | 0.00 | 2.55 | 2.45 | 1 | -1.45 | 0.50 | 0.50 |
| | P | 0.80 | 1.75 | 0.00 | 0.00 | 2.55 | 2.45 | 1 | -1.45 | 0.50 | 0.50 |

Information about the band type, band overlap population (BOP), band length, total number of each type of bond of the $AlPO_4$ has been obtained (we have not shown the extended set of data here). High value of BOP represents the strong covalent bond while the low values represent the ionicity of chemical bonding. We have found that in $AlPO_4$, the bond O-P has stronger covalent character than the O-Al bonding. The weakest bond is formed between the O-O atoms. The bond lengths for O-P, O-Al, and O-O are approximately 1.502, 1.704, and 2.448 Å, respectively.

The estimated Mulliken bond population represents the electron density linked to a bonding. The zero (or close to zero) overlap population indicates the absence of bonding between the electronic populations of the two atoms. The positive bond overlap populations indicate the bonding interactions between the atoms involved [21]. The negative overlap populations indicate the anti-bonding states or electrostatic repulsion between the atoms. $AlPO_4$ have both bonding- and anti-bonding-type interactions. We found 54 bonds in the unit cell; the anti-bonding states are from the negative populations of O-O bonds. The BOP between the O-O atoms is – 0.20. On the other hand, the BOP for the O-P atoms is +0.65, the highest among all the bonds. Large numbers of anti-bonding states lower the hardness of a solid [44,45].

### *3.5. Charge Density Difference (CDD) Maps*

For deeper understanding of inter-atomic bonding nature in hexagonal $AlPO_4$, we have studied the electronic charge density difference (CDD) within the crystal structure. It shows the accumulation and depletion of charges around different atoms in the compound. Covalent bonding has charge accumulation between the atoms involved. For ionic bonding there is charge accumulation around one of the atoms and charge depletion around the other. Fig. 6 shows the CDD profile. On the right hand side of the figure, color scale shows the electronic charge density difference (in e/Å$^3$). The red and blue colors indicate the high and low value of electronic charge density, respectively. It is seen that close to the Al and P atoms there are charge reduction (blue region) but there is significant charge gathering (yellow region) around the O atoms. Thus, electronic charges are transferred from Al and P atoms to the O atoms. This is indicative of ionic bonding. There is some charge accumulation between P-P atoms in the (111) plane which implies weak covalent bonding. In the (001) plane, charge accumulation is observed in between the Al-P and O-P atoms, indicative of covalent bondings. Therefore, the overall bonding in $AlPO_4$ is expected to be a mixture of both ionic and covalent. The maximum electron density is observed around the O atoms in both the planes compared to the P and Al atoms.

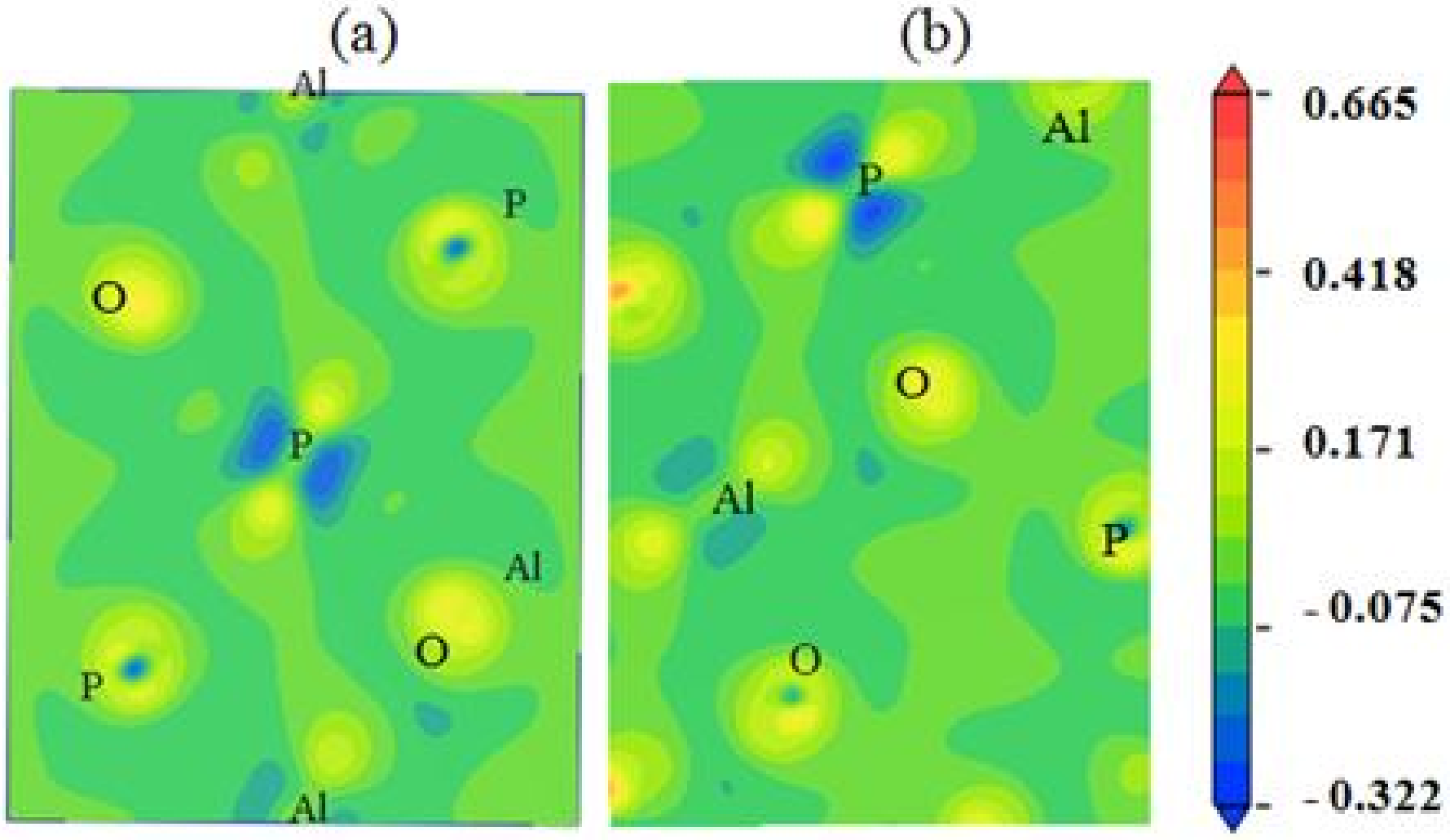

Figure 6: The electronic charge density distribution (CDD) map for $AlPO_4$ in the (a) (111) and (b) (001) planes. The color scale on the right quantifies the amount of charge (in unit of electronic charge).

*3.6. Optical parameters*

The study of optical properties is important to judge the prospect of a material for possible optoelectronic applications [31]. In this section we have shown the energy and electric field polarization dependent optical parameters of hexagonal $AlPO_4$. We have calculated dielectric constant $\varepsilon(\omega)$, refractive index n(ω), absorption coefficient α(ω), photoconductivity σ(ω), reflectivity R(ω), and loss function L(ω) of $AlPO_4$. The optical parameters are calculated for photon energies up to 15 eV for two electric field polarization vectors in the [100] and [001] directions. A Gaussian smearing of 0.5 eV is used for all the calculations [46].

The real part of dielectric constant $\varepsilon_1(\omega)$ is related to the electric polarization. The imaginary part, $\varepsilon_2(\omega)$ is related to the dielectric loss in the medium. The dielectric function is shown in Figure 7(a). The peaks in $\varepsilon_1(\omega)$ are observed at ~9.11 eV and 8.33 eV for the [100] and [001] polarizations and the peaks in $\varepsilon_2(\omega)$ are observed at ~10.2 eV and 9.91 eV for the [100] and [001] polarizations, respectively. The real part of the dielectric constant at zero photon frequency gives the static dielectric constant; the value of which is 2.29. Both real and imaginary parts show anisotropy in the ultraviolet energy region. The overall optical anisotropy in the dielectric function is moderate.

The complex refractive index of a material is expressed as: $N(\omega) = n(\omega) + ik(\omega)$, where $k(\omega)$ is the extinction coefficient. The energy dependent behavior of $n(\omega)$ is shown in Figure 7(b). This parameter determines the phase velocity of light in the medium. The imaginary part of refractive index $k(\omega)$ determines the conductive loss in the material. The energy dependent refractive index $k(\omega)$ of $AlPO_4$ is shown in Figure 7(b). The refractive index in the low energy ultraviolet (UV) region is quite high. This makes $AlPO_4$ a suitable compound for wave guide and photonic engineering devices operating in the UV region. $AlPO_4$ has completely transparent properties in visible light region due to its large insulating bandgap.

The optical absorption coefficient $\alpha(\omega)$ of the $AlPO_4$ as a function of photon energy is shown in Figure 7(c). The absorption begins from ~6.0 eV photon energy for both the polarization directions. This energy threshold matches well with the band gap value obtained from the electronic band structure and DOS profile. It is seen that the compound under study is a very efficient absorber of mid-UV light. The absorption coefficient increases in the UV region and reaches peak at around 13.0 eV and 12.7 eV for [100] and [001] polarization directions, respectively. A sharp reduction of absorption coefficient happens at energy around 14 eV. Material of high absorption coefficient is widely used in optoelectronic devices in both visible and ultraviolet regions. Hence, $AlPO_4$ compound is promising candidates for optoelectronic devices.

The optical conductivity, $\sigma(\omega)$ of a material is an important tool which signifies its electrical conductivity in the presence of an alternating electric field. For $AlPO_4$ the photoconductivity starts at 6.07eV photon energy (Figure 7(d)) as expected for a large band gap insulator. Optical conductivity becomes anisotropic above 9.5 eV. The maximum photoconductivity of the compound is found at 10.4 eV and 10.1 eV for [100] and [001] polarizations, respectively in UV region.

The reflectivity, $R(\omega)$ of a material is a necessary optical parameter which is the ratio of the reflected to incident photon energy. It also inform about the electronic structure of material. The reflectivity curves show that it starts with a value of 4.13 % and rises to maximum values of 16.4% for $AlPO_4$ which is shown in Figure 7(e). The reflectivity curve for our compound is almost non-selective in the visible light region and rises sharply in the UV region. Very low value of reflectivity of hexagonal $AlPO_4$ in the visible region suggests that this compound can be used as an efficient anti-reflection coating to reduce solar heating.

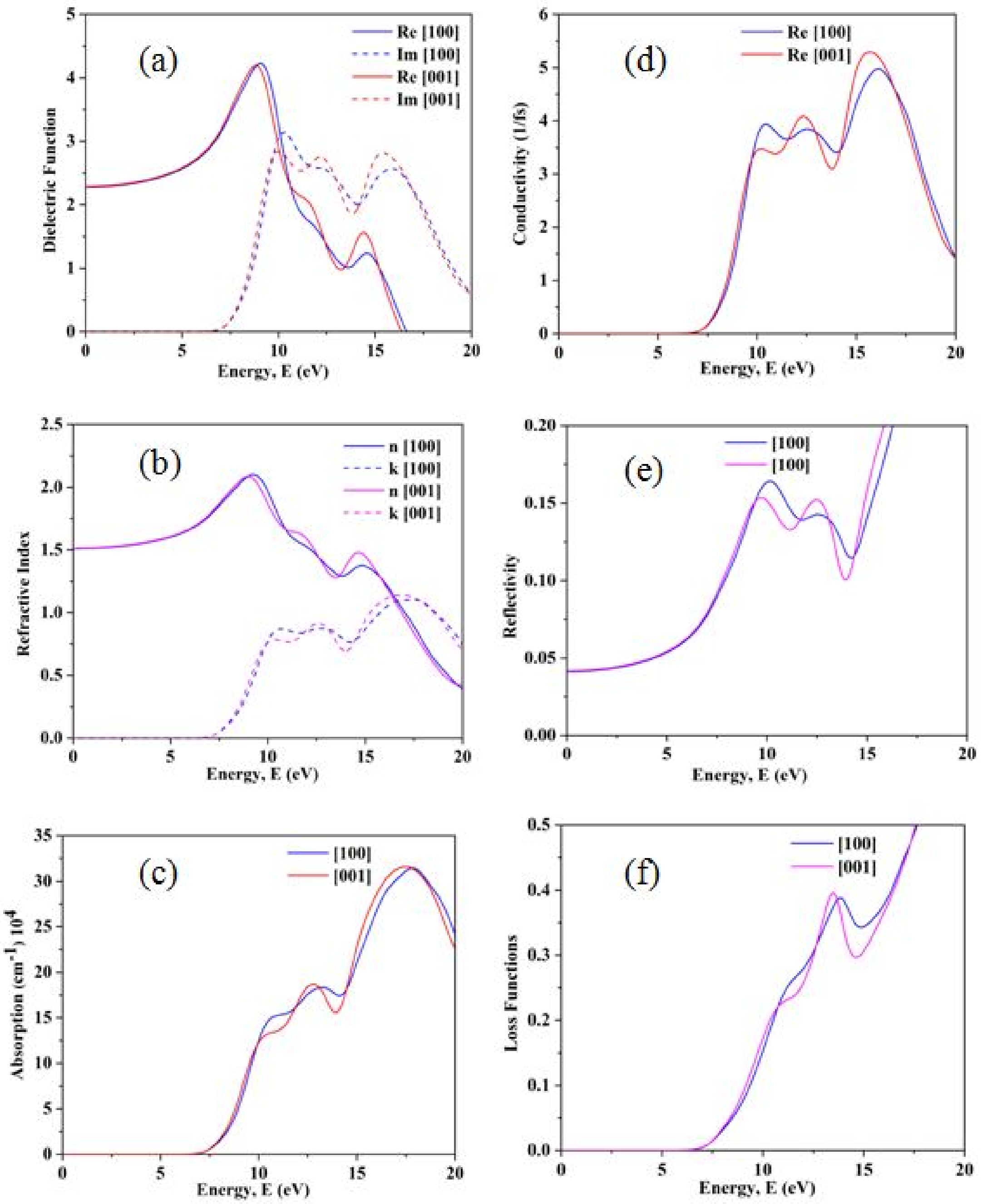

Figure 7: Energy dependent (a) real and imaginary parts of dielectric function [$\epsilon_1(\omega)$ and $\epsilon_2(\omega)$], (b) real and imaginary parts of the refractive index [n(ω) and k(ω)], (c) absorption coefficient [α(ω)], (d) optical conductivity [σ(ω)], (e) reflectivity [R(ω)] and (f) loss function [L(ω)] of $AlPO_4$ for the [100] and [001] electric field polarization directions.

The electron energy loss spectrum as a function of frequency for $AlPO_4$ is shown in Figure 7(f). The plasma frequency ($\omega_p$) can be obtained from the loss energy spectrum since the peak in L(ω) is associated with the plasma resonance and the corresponding frequency is called the plasma frequency ($\omega_P$) [47]. A weak peak in the loss function is found around 13.8 eV for both polarizations of the incident electric field. Both reflectivity and absorption coefficient fall sharply around this particular energy.

*3.7. Thermo-physical Parameters*

The Debye temperature is one major physical parameter in solid-state physics. It is related to the lattice dynamics, phonon heat capacity, and phonon thermal conductivity. The strength of atomic bonding and hardness of a solid can also be gauged from its Debye temperature. Theoretically, at Debye temperature, all the atomic modes of vibrations are activated. Thus, Debye energy represents the cutoff energy for the phonon modes. The high-frequency modes have to be considered as frozen below the Debye temperature ($\theta_D$). The Debye temperature of $AlPO_4$ is calculated from the elastic constants, by using the following formula proposed by Anderson [48]:

$$\theta_D = \frac{h}{K_B}\left[\left(\frac{3n}{4\pi v_0}\right)\right]^{\frac{1}{3}} V_m \quad (44)$$

with,

$$V_m = \left[\frac{1}{3}\left(\frac{1}{V_l^3} + \frac{2}{V_t^3}\right)\right]^{-\frac{1}{3}} \quad (45)$$

$$V_l = \left[\frac{3B+4G}{3\rho}\right]^{\frac{1}{2}} \quad (46)$$

$$V_t = \sqrt{\left(\frac{G}{\rho}\right)} \quad (47)$$

Here, h is Planck's constant, $k_B$ is the Boltzmann's constant, $V_0$ refers to the volume of unit cell, $V_m$ is the average sound velocity through the solid, n defines the number of atoms in the cell and $V_t$ and $V_l$ are the transverse and longitudinal sound velocities through the solid, respectively [49].

Calculated Debye temperature of $AlPO_4$ is 356.2 K which is listed in Table 10.

Table 10: Number of atoms per unit volume n (atoms/m$^3$), Debye temperature $\Theta_D$ (K), melting temperature $T_m$ (K), thermal expansion coefficient α ($K^{-1}$), the lattice thermal conductivity $K_{ph}$ (W/m.K) at 300 K, Grüneisen parameter γ, minimum thermal conductivity $K_{min}$ in Cahill's and Clark's method ( W/m.K) of $AlPO_4$.

| Compound | n ($\times 10^{28}$) | $\Theta_D$ | $T_m$ | α ($\times 10^{-5}$) | $k_{Ph}$ | γ | $K_{min}$ | | Ref. |
|---|---|---|---|---|---|---|---|---|---|
| | | | | | | | Cahill | Clark | |
| $AlPO_4$ | 2.589 | 356.20 | 667.63 | 4.28 | 16.85 | 0.784 | 0.618 | 0.488 | This work |

Electrons and phonons both transport thermal energy through the solids. In insulators, the phonon contribution dominates. The lattice/phonon thermal conductivity, $k_{ph}$ is major thermo-physical parameter of solids on which various technical applications like as in heat sinks, sensors, development of new thermoelectric materials, transducers and thermal barrier coatings [33,50] depend. Thermoelectric devices, solid state refrigeration and thermal barrier coating (TBC) systems require low thermal conductivity materials [50]. On the other hand, high thermal conductivity materials with minimum heat waste are important to improving the efficiency of heat remover in microelectronic and nanoelectronic devices. We have calculated Phonon thermal conductivity, $k_{ph}$, of $AlPO_4$ at 300 K using the formula put forward by Slack, as follows [51]:

$$k_{Ph} = A(\gamma)[\frac{M_{av}\theta_D^3\delta}{\gamma^2 N^{\frac{2}{3}} T}] \quad (48)$$

$$\gamma = \frac{3[1+\sigma]}{2[2-3\sigma]} \quad (49)$$

Here, $M_{av}$ is the average atomic mass per atom in the compound (kg/atom), $\theta_D$ is the Debye temperature (K), δ refers to the cubic root of average atomic volume (m), N represents the total number of atoms present in the unit cell, T defines the absolute temperature (K), σ is the Poisson's ratio and γ is the Grüneisen parameter that measures the anharmonicity of phonons. In general, phonon thermal conductivity is low in materials with high γ. We can estimate the coefficient A(γ) in Eq.(48) as follows [34,52]:

$$A(\gamma) = \frac{5.720\times10^7\times0.849}{2\times[1-\frac{0.514}{\gamma}+\frac{0.224}{\gamma^2}]} \quad (50)$$

The calculated phonon thermal conductivity at room temperature of $AlPO_4$ is listed in Table 10. The low value of $k_{ph}$ indicates weak covalent bonding. It also results from soft phonon modes [53]. The theoretical lower limit of intrinsic lattice thermal conductivity can be found using the modified Clarke's model [26,54,55]:

$$K_{min}^{Clarke} = k_B V_m [V_{atomic}]^{-\frac{2}{3}} \quad (51)$$

where, $V_{atomic}$ is the average volume per atom, $V_m$ is the average sound velocity in solid. Cahill [55] also proposed the following expression to calculate the minimum thermal conductivity:

$$K_{min}^{Cahill} = \frac{k_B}{2.48} n^{\frac{2}{3}} ( V_l + 2V_t ) \quad (52)$$

where, n is the number of atoms per unit volume, $V_l$ and $V_t$ are the longitudinal and transverse sound velocities in the solid. The calculated minimum thermal conductivity, $K_{min}^{Clarke}$, $K_{min}^{Cahill}$ and the Grüneisen parameter, γ are given in Table 10. It has been reported earlier that ionic bonding and high Grüneisen parameters can lead to low thermal conductivity [24,51].

The melting temperature, $T_m$ is also an important property of a material for thermal management. The estimated value of $T_m$ for $AlPO_4$ in the ground state is 667.63 K (Table 10). Low $T_m$ materials are good candidates as thermal interface. Below $T_m$ solids are thermodynamically stable and can operate continuously without chemical change, oxidation and extra distortion leading to mechanical failure. The melting temperature $T_m$ increases with the bonding strength, higher cohesive energy and lower coefficient of thermal expansion [50]. The $T_m$ of hexagonal $AlPO_4$ has been calculated using following expression [26,56]:

$$T_m = 354K + \frac{1.5K}{GPa}(2C_{11} + C_{33}) \quad (53)$$

The thermal expansion coefficient (TEC) is another important thermal property of materials. Low thermal expansion materials are used in the ceramic industry [31,57,58]. The TEC can be calculated from the shear modulus, G (in GPa) of a material by using the following empirical equation [50]:

$$\alpha = \frac{1.6\times10^{-3}}{G} \tag{54}$$

The calculated value of α for $AlPO_4$ at 300 K is given in Table 10.

Last few years many ceramic materials, usually oxides have been suggested as new thermal barrier coating (TBC) materials. TBC materials having duplex-type of coat like metallic bond coat and a ceramic topcoat. This coat protects the substrates from oxidative and harmful attack with developed the bonding between ceramic topcoat and materials [59,60]. Efficient TBC systems are characterized by low thermal conductivity, low TEC and high melting temperature. High melting temperature is required when the system is designed to operate at high temperatures. $AlPO_4$ in the hexagonal structure has very low lattice thermal conductivity. Since it is a wide band gap insulator, electronic contribution to the thermal conductivity is expected to be very low. The TEC of this compound is also moderate. The melting temperature, on the other hand, is low as well. Thus, this compound has a potential to be used as a TBC material at low temperatures.

## 4. Conclusions

We have presented a thorough investigation of the physical properties of hexagonal $AlPO_4$ by employing the first-principles DFT calculations. The mechanical properties, elastic anisotropy, electronic band structure, bonding properties, optical properties, and thermo-physical properties have been explored. Most of the results obtained are novel. The system under study is a layered, elastically stable, brittle compound. Significant ionic and covalent bondings are present. The compound is soft in nature. Electronic band structure calculations show that hexagonal $AlPO_4$ is a wide bandgap insulator. SOC has only a minor effect on the band structure. Soft nature and very large TDOS just below the Fermi level suggests that pressure can be used quite efficiently to tune the electronic ground state of this compound. The same is true for suitable atomic substitution. The optical parameters show that $AlPO_4$ is a good absorber of UV light. The compound can also be used as anti-reflection coating for visible light to reduce solar heating. It has some potential to be used as a TBC at low temperatures. The refractive index of $AlPO_4$ in the visible and low energy ultraviolet region is quite high. For this, it can be useful for wave guide and photonic engineering devices.

In summary, we have investigated systematically a large number of hitherto unexplored physical properties of hexagonal $AlPO_4$ in this paper. We hope that these results will stimulate researchers to investigate the compound in further detail, both theoretically and experimentally in near future.

## Acknowledgements

S. H. N. acknowledges the research grant (1151/5/52/RU/Science-07/19-20) from the Faculty of Science, University of Rajshahi, Bangladesh, which partly supported this work. A. S. M. M. R. is thankful to the University Grant Commission (UGC), Dhaka, Bangladesh which gave him the Fellowship to conduct this research.

## Data availability

The data sets generated and/or analyzed in this study are available from the corresponding author on reasonable request.

## Declaration of interest

The authors declare that they have no known competing financial interests or personal relationships that could have appeared to influence the work reported in this paper.

## CRediT authorship contribution statement

**A.S.M. Muhasin Reza**: Formal analysis, Methodology, Writing–original draft. **Md. Asif Afzal:** Formal analysis, Methodology. **S.H. Naqib**: Supervision, Formal analysis, Conceptualization, Project administration, Writing-review & editing.